# $Cr^{3+}$ spin dynamics under the octahedral crystal field in van der Waals antiferromagnets

Rabindra Basnet,[1] Subhashree Chatterjee,[1] Paul Kigaya,[1] Ezana Negusse,[1] J. van Tol,[2] and Ramesh C. Budhani[1,*]

[1]*Department of Physics, Morgan State University, Baltimore, MD, 21251, USA*
[2]*National High Magnetic Field Laboratory, Florida State University, Tallahassee, Florida 32310, USA*

## Abstract

The magnetic moment in van der Waals (vdW) materials containing 3*d* transition metals originates from unpaired *d*–electron spins and their interaction with surrounding ligands. The interplay between exchange interactions and magnetic anisotropy stabilizes long–range ordering of such moments. The compound $CuCrP_2S_6$ (CCPS) presents an interesting class of vdW solids where the coupling of $Cr^{3+}$ moments and ordering of $Cu^{1+}$ ions give rise to a multiferroic ground state. Here we investigate the spin dynamics of $Cr^{3+}$ ions in CCPS through magnetization and broadband as well as single sub–THz magnetic resonance measurements. The orbital moment of $Cr^{3+}$ is quenched under the octahedral crystal field of surrounding chalcogen ions, resulting in negligible magnetic anisotropy—a feature common to Cr–based vdW antiferromagnets (AFM). Resonance spectra over a wide frequency–field–temperature range reveal quasi-2D AFM dynamics governed mainly by isotropic Cr–Cr exchange interactions, which determine the magnetic order, spin reorientation, and damping. Sub–THz resonance spectra also uncover a field–induced ferromagnetic polarization, highlighting the universal role of Cr–Cr exchange in layered Cr compounds. Moreover, persistent magnetic correlations far above the Néel temperature ($T_N \approx 32$ K) points to short–range magnetic order in CCPS and motivates future studies of a possible interplay between AFM and antiferroelectric orders. These results establish CCPS as an exemplary system for exploring 2D magnetism and electric-field-tunable spintronic functionalities in layered multiferroics.

[*]ramesh.budhani@morgan.edu

## I. INTRODUCTION

Magnetic materials are constructed from magnetic moments located on the atomic sites of a crystal lattice, where they interact through exchange interactions of Coulombic and dipolar character. These moments originate predominantly from the unpaired $d$– or $f$–electron spins of the ions, which determine their total moment as well as the nature and strength of exchange and anisotropy. Van der Waals (vdW) magnets have recently emerged as key platforms for studying low–dimensional magnetism and its technological applications [1,2]. Amongst the $3d$ metal cations in vdW magnets, materials based on $Cr^{3+}$ ions have garnered significant attention due to their robust magnetic ordering from bulk to the monolayer limit [3–6]. Compounds such as $CrI_3$ [3] and $Cr_2Ge_2Te_6$ [4] were the first experimentally confirmed Cr–based 2D magnets. In general, Cr compounds are of great importance due to the plethora of rich magnetic phases they host. For example, $Cr(Br,I)_3$ [3,7–9] and $Cr_2(Si,Ge)_2Te_6$ [4,10] exhibit ferromagnetism (FM) while $CuCrP_2S_6$ [11–13], $CrPS_4$ [14], $CrCl_3$ [15], and CrSBr [16] are antiferromagnets (AFMs). These compounds possess octahedral ($O_h$) symmetry, with $Cr^{3+}$ ions forming $CrX_6$ octahedra. The $3d^3$ electronic configuration of $Cr^{3+}$ yields a half–filled $t_{2g}$ manifold and empty $e_g$ states, leading to strong orbital–moment quenching. As a result, spin–orbit coupling (SOC) and magnetocrystalline anisotropy are intrinsically weak, a feature often cited as the origin of the flexible magnetic ordering observed in such systems [17–22].

Furthermore, the half–filled $t_{2g}$ configuration influences the nature of magnetic exchange pathways. It usually favors the direct $d$–$d$ hopping, which typically promotes AFM coupling [23]. Nevertheless, the FM ordering also emerges in Cr systems, particularly those with heavier ligands [3,4,7–10]. While the Cr $d$–orbitals contribute similarly across all Cr–based vdW magnets, the key difference stems from the ligand $p$–orbitals, where anisotropic $p$–$d$ superexchange dictates the nature of the magnetic ground state [19–21,24,25]. The magnetic exchange and anisotropy are sensitive to the SOC strength of $p$–electrons and the degree of $p$–$d$ covalency [21], which depend on the chemical identity and spatial geometry of the coordinating ligands. This delicate balance between AFM and FM ground states is a distinctive feature in $Cr^{3+}$ systems; and therefore, unraveling its microscopic origin necessitates further investigations, including the studies of spin dynamics over wider timescales.

Magnetic resonance spectroscopy is an ideal tool for studying magnetization dynamics [9,12,24,26–32]. These include ferromagnetic resonance (FMR) [9,24,27,29,31] and electron paramagnetic/spin resonance (EPR/ESR) [26,27,30,32], which have been used to probe the magnetic order in Cr–based vdW compounds. Such studies demonstrate the 2D magnetic correlations [26,29,30,32] and the presence of multi-domain structures [9,31]. However, the earlier works have been limited to FM and paramagnetic (PM) resonances, primarily due to the high resonance frequencies of AFMs, which typically fall in the terahertz (THz) range, requiring advanced THz spectroscopies to access antiferromagnetic resonance (AFMR) [33,34]. Nonetheless, a few Cr–based magnets, such as $CrCl_3$ [28], $CuCrP_2S_6$ [12], and CrSBr [35], exhibit AFMR at relatively lower frequencies, within the typical microwave range. These studies have revealed strong magnon-magnon coupling in $CrCl_3$ [28] and the multiple magnon modes during the spin-flop (SF) transition in $CuCrP_2S_6$ [12].

In this work, we have studied the spin dynamics of $Cr^{3+}$ ions subjected to an octahedral crystal field in the layered AFM compound $CuCrP_2S_6$ (CCPS) using magnetization and GHz frequency AFMR measurements. CCPS has attracted considerable interest recently for its multiferroic properties and robust magnetoelectric character extending down to the monolayer limit [36,37], making it important for exploring the coupling between magnetic and ferroelectric order parameters in two–dimensions. The temperature and field–dependent susceptibility measurements suggest orbital moment quenching of the $Cr^{3+}$ ion, which is further confirmed by AFMR spectroscopy down to $T = 1.2$ K arising from the quasi–2D AFM dynamics of CCPS. The pronounced resonant absorption of microwaves across a broad frequency, temperature, and field ($f$–$T$–$H$) phase space reveals negligible anisotropy, indicating that the magnetic order, spin reorientation, and damping are primarily governed by the isotropic Cr–Cr exchange interactions. Moreover, the magnetization and high–frequency (240 GHz) FMR measurements uncover a field-induced FM-like spin polarization originating presumably from the intrinsic Cr–Cr interactions. These findings provide fresh insights into the fundamentals of magnetic ordering in vdW antiferromagnets, together with establishing a pathway for microwave control of AFM dynamics. Further, the evidence of room–temperature ferroelectricity in very thin platelets and the exceptionally low damping potentially make CCPS a promising vdW material for next–generation technological applications.

## II. EXPERIMENT

The single crystals of CCPS used in this work were synthesized by chemical vapor transport (CVT) using iodine as the transport agent. First, ~1.2 g of the stoichiometric mixture of Cu (99.9%, Thermo Scientific), Cr (99.97%, Thermo Scientific), P (99%, Beantown Chemical), and S (99.99%, Thermo Scientific) together with 40 mg of $I_2$ (99.8%, Thermo Scientific) was sealed in an evacuated (pressure ~$10^{-5}$ Torr) quartz tube using an oxygen-acetylene torch. As shown in the inset of Fig. 2a, the millimeter-sized single crystals with flat and shiny surfaces were obtained after 1 week of CVT growth with a temperature gradient from 750 to 650°C. These crystals were characterized by X–ray diffraction (XRD) and Raman spectroscopy at room temperature. The XRD was performed using a Rigaku miniflex diffractometer with Cu–$K\alpha$1 radiation. A typical diffraction pattern of the crystals placed such that the scattering vector is parallel to the c-axis has only the (00$L$) reflections (Fig. 2a). The Raman measurement was conducted with an XploRA PLUS confocal Raman microscope (Horiba). The surface topography, thickness, ferroelectricity, and polarization switching of an exfoliated CCPS nanoflake on a Pt–coated Si substrate were studied using an NX10, Park Systems AFM at room temperature. The ferroelectric and piezoelectric characteristics were measured using a Pt/Cr coated conducting tip with a tip sensitivity of 33.33 V/μm and a force constant of 3 N/m with an AC drive voltage of 1 V in off–Resonance piezoresponse force microscopy (OR–PFM) mode (17 kHz), throughout all the measurements. The magnetization measurements were performed in a physical property measurement system (PPMS EverCool). For the AFMR spectroscopy, a custom–built cryo–FMR spectrometer [38], which operates in the frequency modulation mode over a bandwidth of 1 to 22 GHz down to 1.2 K, has been used. The sample was placed on the U-shaped signal line of a grounded co-planar waveguide where the frequency-modulated GHz input from an RF source is fed, and the rectified return signal is measured using lock-in detection. The mutually perpendicular static and RF magnetic fields ($H_{RF}$) were along the sample plane (i.e., $ab$–plane) in these experiments. High–frequency EPR measurements were performed at the National High Magnetic Field Laboratory (NHMFL) located at Florida State University [26]. The sample was cooled down to 5 K under zero–field–cooling conditions to ensure the solidification of the vacuum grease used for mounting the sample. These experiments were conducted over the temperature range of 5 to 260 K, at 240 GHz excitation and a dc magnetic field aligned parallel to the $ab$–plane of the crystal. Measurements were also conducted at different field orientations from out–of–plane $\theta = 0°$ to in-plane $\theta = 90°$ directions at 10 K. The plate–like sample was mounted on a sample rotator, allowing both temperature (at fixed angle) and angular–dependent (at fixed temperature) EPR measurements.

## III. RESULTS AND DISCUSSION

A transition metal ion in free space exhibits full rotational symmetry SO(3), resulting in five–fold degeneracy of its $d$–orbitals. However, this symmetry is lowered in a crystalline environment. In transition metal ($M$) based vdW materials, the $M$ atoms typically form a planar honeycomb network within each layer and coordinate octahedrally with ligand chalcogen atoms ($X$) through edge–sharing octahedra. Therefore, despite distinct crystal structures, the vdW magnets often share a common structural motif, where each $M$ atom is centrally positioned within an $MX_6$ octahedron, coordinated by four ligands in the basal plane and two in the out–of–plane direction, as shown in Fig. 1a. An ideal octahedron features six ligands equidistant from a central $M$ atom. Under such octahedral coordination, the full rotational SO(3) symmetry is reduced to the discrete symmetry group of the octahedra, represented as $O_h$. The crystal electric field created by the $O_h$ symmetry of the transition metal cations lifts the five–fold degeneracy of the $d$–states, splitting it into lower–energy $t_{2g}$ and higher–energy $e_g$ orbitals. Among the five $d$–orbitals, the $d_{x^2-y^2}$ and $d_{z^2}$ orbitals point directly towards the surrounding ligands (Fig. 1a). Consequently, the electrons in these orbitals experience greater electrostatic repulsion, pushing them to a higher energy level and creating a two-fold degenerate high-energy $e_g$ orbital. On the other hand, the remaining three $d$–orbitals $d_{xz}$, $d_{yz}$, and $d_{xy}$ do not directly align towards ligands (Fig. 1a), thus experiencing a weaker repulsion. This group of orbitals, commonly referred to as $t_{2g}$, which is three–fold degenerate, lies below the $e_g$ orbitals [23,39]. This splitting of the five $3d$ orbitals under the octahedral crystal field is depicted in Fig. 1a. Table I highlights the fact that most Cr–based vdW magnets exhibit $O_h$ symmetry with $Cr^{3+}$ ion forming a $CrX_6$ octahedron with the surrounding ligands (Fig. 1a; left panel). In certain cases, such as in $CrSiTe_3$ [40] and CrSBr [16], the octahedron is slightly distorted, where the out–of–plane ligands are often located at different distances from the central $M$ atom than the in–plane ligands [23]. This further lowers the local point symmetry to $D_{3d}$ (trigonal distortion) and $D_{2v}$ (orthorhombic distortion), respectively. As mentioned earlier, the half–filled $t_{2g}$ orbital for the $Cr^{3+}$ ion under $O_h$ symmetry typically favors AFM ordering [23]. However, in systems with heavier ligands, the enhanced covalency due to greater orbital overlap facilitates FM superexchange interactions, ultimately stabilizing the FM ground state [19–21]. Hence, as presented in Table I, Cr compounds such as CCPS [11–13], $(Ni_{0.91}Cr_{0.09})PS_3$ [41], $CrPS_4$ [14], and $CrCl_3$ [15] containing lighter ligands exhibit an AFM ground state. In contrast, FM ordering emerges in systems with heavier ligands like Te, Br, and I. Even in AFM Cr compounds, the signatures of ferromagnetism are still evident. This delicate balance between the AFM and FM ground states, depending on the nature of the chalcogen, suggests a high propensity of the AFM state to a magnetic field–induced spin flop (SF) transition to the FM state. A conceptual schematic of this field–induced magnetic phase transition is depicted in Fig. 1b. However, this picture needs experimental verification and an in-depth analysis to uncover the mechanism responsible for such a phase transition in the $Cr^{3+}$–based vdW magnets.

*Table I. Magnetic properties of a few well-known Cr-based vdW antiferromagnets. In these materials, the $Cr^{3+}$ ion is subjected to an octahedral ($O_h$) or distorted octahedral ($D_{2v}$ or $D_{3d}$) crystal electric field arising from the coordination of surrounding ligands. Here, IP and OOP denote in–plane and out–of–plane directions, respectively.*

| Materials | Crystal field symmetry | Magnetic structure | $T_N/T_C$ (K) | $B_{SF}$ (T) | $B_S$ (T) | Ref. |
|---|---|---|---|---|---|---|
| $CuCrP_2S_6$ |  | IP A-type AFM | ~32 | ~0.4 | ~6.4 | This work |
| $(Ni_{0.91}Cr_{0.09})PS_3$ |  | IP AFM | ~34 | ~0.6 | ~8.0 | [41] |
| $CrPS_4$ |  | OOP A-type AFM | ~38 | ~0.7 | ~8.0 | [14] |
| $CrCl_3$ | $O_h$ | IP A-type AFM | ~14 | ~0.015 | ~0.25 | [15] |
| $CrBr_3$ |  | OOP FM | ~37 | NA | ~0.20 | [7,9] |
| $CrI_3$ |  | OOP FM | ~61 | NA | ~0.20 | [8,9] |
| $CrGeTe_3$ |  | OOP FM | ~67 | NA | ~0.25 | [61] |
| $CrSiTe_3$ | $D_{3d}$ | OOP FM | ~33 | NA | ~0.40 | [10,40,62] |
| $CrSBr$ | $D_{2v}$ | IP A-type AFM | ~132 | ~0.3 | ~0.58 | [16] |

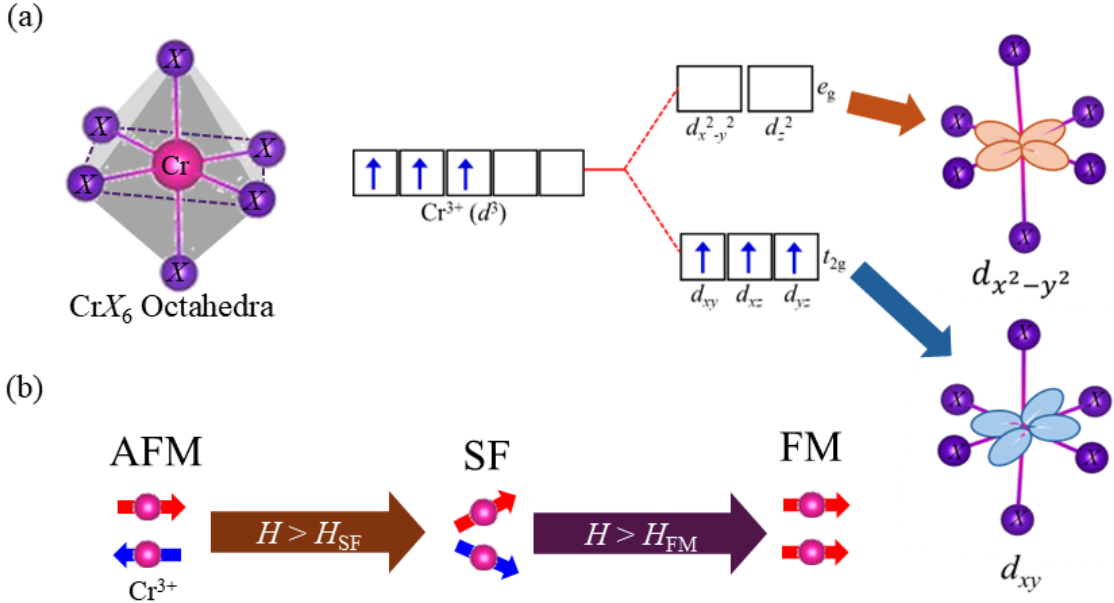

*Figure 1. (a) Left panel: CrX$_6$ octahedron, where the pink and purple spheres denote Cr and X atoms. Here, X represents ligands such as chalcogen, halogen, etc. Right panel: Splitting of five 3d orbitals of Cr$^{3+}$ cation into three t$_{2g}$ and two e$_g$ levels under an octahedral crystal field. The three 3d electrons of the Cr$^{3+}$ ion occupy the lower t$_{2g}$ orbitals, resulting in a half–filled configuration. The schematics also show the d$_{xy}$ and d$_{x^2-y^2}$ orbitals with respect to the positions of the ligand atoms. (b) Conceptual schematic of the field–driven transition from AFM to SF state, followed by FM–like polarization of the Cr$^{3+}$ moments.*

With this motivation, we investigated the magnetic behavior of Cr$^{3+}$ spin in vdW antiferromagnets, using CCPS as a representative material, which belongs to the metal thiophosphate (MPX$_3$) family of compounds. It is formed by replacing the bivalent metal ion $M^{2+}$ with an equimolar ratio of monovalent Cu$^{1+}$ and trivalent Cr$^{3+}$ ions. Incorporating heterovalent metal ions in the formula adds chemical and structural complexity while maintaining charge neutrality in the system. Clearly, careful control of experimental conditions is essential to ensure successful crystal growth of this compound [11,42]. Fig. 2a (Inset) presents the optical microscope image of a millimeter-sized CVT-grown single crystal of CCPS, which exhibits a high-intensity (00L) XRD spectrum (Fig. 2a), consistent with the reported monoclinic (C2/c space group) crystal structure [42]. Furthermore, room-temperature Raman measurement identifies four prominent modes, labeled I, II, III, and IV (Fig. 2b), that agree well with earlier Raman studies on CCPS [43,44]. These four peaks are assigned to the vibrational modes of Cu$^{1+}$ and Cr$^{3+}$ ions as well as symmetric stretching and deformation of PS$_3$ units within the (P$_2$S$_6$)$^{4-}$ anions [44]. In CCPS, while the Cr$^{3+}$ ions reside at the center of each sandwich layer within the vdW structure, the Cu$^{1+}$ ions are off–centered along the *c*–axis [11,13,44,45] (Fig. 2c, left figure), breaking the inversion symmetry of the lattice and thereby inducing a spontaneous electric polarization in each layer even at ambient temperature [36]. To confirm the room-temperature ferroelectricity, we performed the PFM measurements on a thin (~48 nm) exfoliated CCPS flake shown in Fig. 2d. The corresponding PFM phase (Fig. 2d; upper panel) and amplitude (Fig. 2d; lower panel) images reveal clear ferroelectric domain contrast across the flake, indicating the presence of spontaneous polarization. To further probe the switchable polarization, a box–in–box pattern was written by applying a ±10 V DC bias, as shown in the phase (Fig. 2e; upper panel) and amplitude (Fig. 2e;

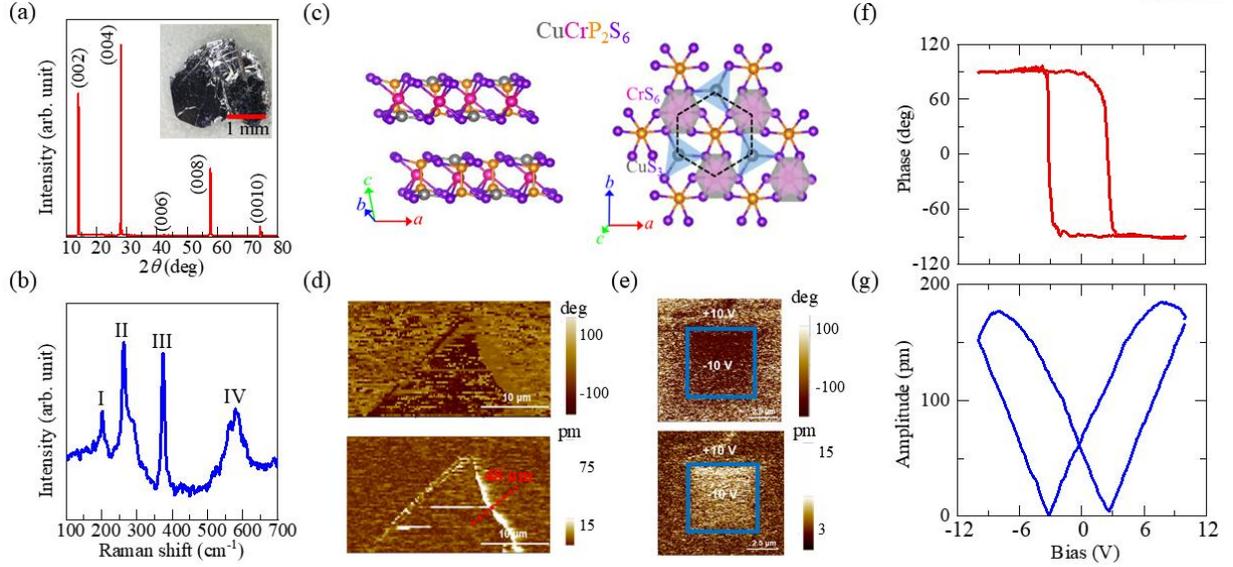

***Figure 2.*** *Growth and characterization of CCPS single crystal. (a) X-ray diffraction pattern of the CCPS crystal. Inset: The optical microscope image of a single crystal. (b) Raman spectra of the CCPS crystal at room temperature, showing four prominent peaks, labeled as I, II, III, and IV. (c) Left panel: The monoclinic crystal structure of CCPS showing a, b, and c axes. Right panel: The arrangement of Cu$^{1+}$ and Cr$^{3+}$ ions in a honeycomb network in the ab–plane (shown by the dashed hexagon), forming quasi-trigonal CuS$_3$ (blue triangles) and octahedral CrS$_6$ (magenta hexagon) units coordinated by S$^{2-}$ ions. (d) PFM phase (upper panel) and amplitude (lower panel) images of a 48 nm–thick CCPS flake obtained by OR-PFM. (e) PFM phase (upper panel) and amplitude (lower panel) images acquired after writing a box–in–box pattern using a ±10 V DC bias. (f) Off–field PFM phase hysteresis loop and (g) corresponding amplitude loop recorded during the DC bias switching process.*

lower panel) maps. The inner and outer regions exhibit opposite phase contrasts, confirming reversible domain switching under applied DC electric fields. The local ferroelectric switching behavior was also shown by measuring the off–field PFM phase hysteresis loop (Fig. 2f), which exhibits a 180° phase shift, a hallmark of ferroelectric polarization switching. The corresponding butterfly–shaped amplitude loop (Fig. 2g) displays the characteristic piezo–response of CCPS, with minima at the coercive voltages where polarization reversal occurs. These results collectively confirm the robust room–temperature ferroelectricity and switchable polarization in the CCPS flake.

Fig. 2c (right figure) illustrates the arrangement of $Cu^{1+}$ and $Cr^{3+}$ ions in a honeycomb network in the *ab*–plane (shown by the dashed hexagon), forming quasi–trigonal $CuS_3$ (blue triangles) and octahedral $CrS_6$ (magenta hexagon) units coordinated by $S^{2-}$ ions. As discussed above, such an $O_h$ symmetry of the $Cr^{3+}$ ion has a significant impact on magnetism. To investigate the magnetic properties of CCPS, we have measured the temperature dependence of susceptibility ($\chi$) under in-plane ($H // ab$) and out-of-plane ($H \perp ab$) magnetic fields of $\mu_0 H = 0.1$ T (Fig. 3a). The susceptibility of the CCPS single crystal exhibits a sharp PM to AFM transition with a Néel temperature $T_N$ of ~32 K (black arrow, Fig. 3a), consistent with the earlier works [11–13,42]. The $T_N$ peak is more pronounced in the in–plane susceptibility ($\chi_{//}$) compared to the out–of–plane ($\chi_\perp$), and $\chi_{//}$ remains smaller than $\chi_\perp$ below $T_N$, indicating the in–plane easy axis in this compound. In fact, CCPS is found to possess an in–plane A–type AFM structure [12,46], with the Cr moments aligned ferromagnetically along the *a*–axis within each layer but forming AFM coupling between the adjacent layers (Fig. 3a, inset). Although this system adapts an AFM ground state, the magnetic order is primarily governed by intra–layer FM interactions, while inter–layer AFM interactions are rather weak [47]. This is supported by the positive Weiss temperature ($\Theta$) obtained from the Curie–Weiss (CW) fitting of the susceptibility data, contrary to the negative $\Theta$ expected for AFM order. As shown by the red fitted line in Fig. 3b, $\chi$ in the PM state can be described by the CW law $\chi = \chi_0 + C/(T - \Theta)$, where $\chi_0$ is the temperature–independent part of susceptibility and $C$ is the Curie constant. The CW fitting yielded a value of $\Theta \approx (28.3\pm0.4)$ K, which is also confirmed by $1/\chi$ vs $T$ plot (right vertical axis, Fig. 3b) and close to the reported values of $\Theta \approx 31.5$ K [46] and $(35\pm1)$ K [11], implying the dominant FM interactions in this material. Additionally, the CW coefficient $C$ from the fit is $\approx (1.86\pm0.1)$ emu K/ (Oe mol), which is used to estimate the effective moment ($\mu_{eff}$) of the system using the relation $\mu_{eff} = \sqrt{3K_B C/N_A}$.

The strong FM correlation within the A–type AFM structure is further illustrated in the field dependence of magnetization [$M$(H)] measurements under $H \parallel ab$ (upper panel) and $H \perp ab$ (lower panel) magnetic fields presented in Fig. 3c. Given the absence of magnetic hysteresis around zero field and the symmetric magnetization response in negative and positive field directions, we have only shown the $M$(H) data from $\mu_0 H = 0$ to + 9 T at all temperatures. CCPS exhibits a metamagnetic SF transition at an in-plane magnetic field of $\mu_0 H_{SF} \approx (0.44 \pm 0.06)$ T at $T = 5$ K, but is absent in the out-of-plane field direction, which is manifested as a clear upturn (shown by a red arrow in Fig. 3d) in isothermal $M$(H) below $T_N$. The Cr–based vdW compounds undergo the SF transition at a relatively low field, typically below $\mu_0 H_{SF} < 1$ T (Table I). This behavior can be interpreted as the emergence of weak FM components, arising from the reorientation of spins from an AFM arrangement into a canted configuration when a field is applied along the easy axis. Applying a field exceeding $H_{SF}$ spontaneously rotates the canted moments towards the field direction and eventually polarizes them above a saturation field ($H_{FM}$). This process is marked by the onset of field–independent magnetization near a saturation field of $\mu_0 H_{FM} \sim (6.4 \pm 0.4)$ T (blue arrow) and $\sim (7.8 \pm 0.4)$ T (purple arrow) in in–plane and out–of–plane magnetizations, respectively, at 5 K. Such a field–driven AFM to SF transition, followed by FM–like polarization at low field of just a few tesla, as depicted in Fig. 1b, appears to be a generic feature in AFM Cr compounds (Table I). The field–driven magnetic phase transition in CCPS is corroborated by the magnetic phase diagram shown in Fig. 3e, which summarizes the $H_{SF}$ and $H_{FM}$ critical fields extracted from the slope changes and kinks in the in–plane magnetization (easy axis), corresponding to the peak (red arrow) and sharp drop (blue arrow) in the derivative d$M$/d$H$ (Inset, Fig. 3c). It also captures their temperature dependence below $T_N$, providing a comprehensive picture of the magnetic states over a wide field–temperature phase space in CCPS. Both the $H_{SF}$ and $H_{FM}$ fields reduce on approaching the $T_N$. The reduction of $H_{FM}$ with the rising temperature is understandable. Here, thermal energy assists with moment reorientation, enabling polarization to occur at a lower field upon heating. On the other hand, the temperature dependence of $H_{SF}$ is intriguing. While the suppression of $H_{SF}$ has been seen in other Cr-based antiferromagnets, such as CrPS$_4$ [14] and CrSBr [16], this is in stark contrast to the opposite trend reported in other AFM compounds of this material family, such as MnPS$_3$ [48] and NiPS$_3$ [49], where it rises with temperature. The SF transition in these compounds is primarily governed by magnetic anisotropy rather than exchange interactions [49]. Therefore, the enhancement of $H_{SF}$ with temperature may originate from the temperature–induced modification of magnetic anisotropy. However, as mentioned earlier, the single-ion anisotropy for the $Cr^{3+}$ ion is negligible. Thus, the moment rotation and the resulting SF transition in Cr–based AFM compounds are likely driven by a distinct mechanism, which will be discussed later in this work.

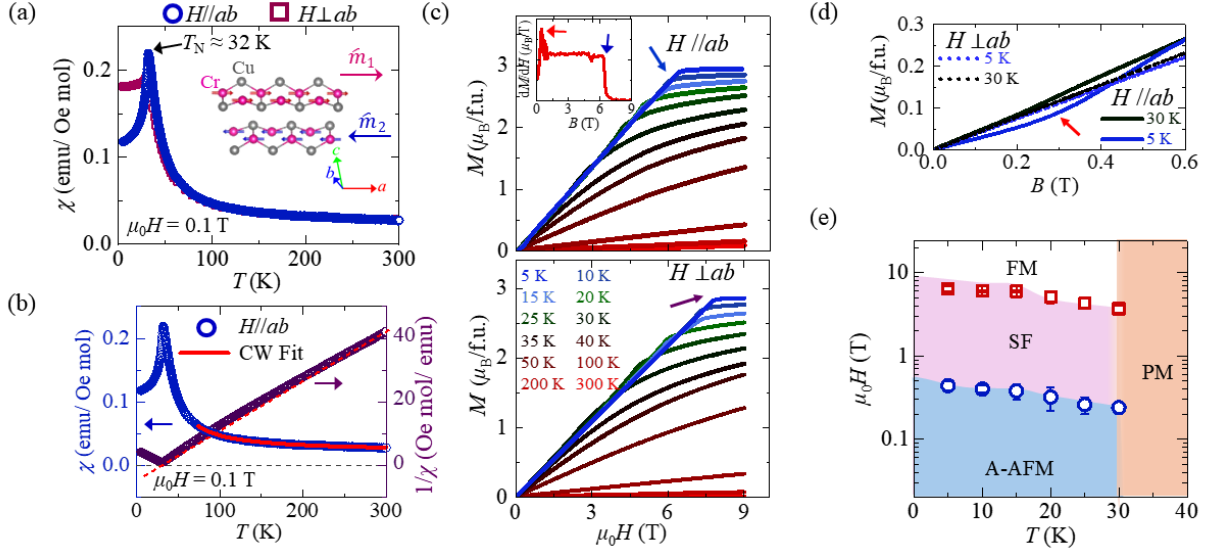

***Figure 3.*** *Magnetic properties of CCPS. (a) Temperature dependence of susceptibility ($\chi$) for the sample under H||ab and H⊥ab fields of $\mu_0H$ = 0.1 T. The black arrow denotes the PM to AFM transition temperature $T_N \approx 32$ K. Inset: The magnetic structure of CCPS, where two AFM sublattices are represented by $\hat{m}_1$ and $\hat{m}_2$. (b) Left vertical axis: The in–plane susceptibility data, reproduced from Fig. 3a, is shown with the Curie–Weiss (CW) fitting represented by the red curve. Right vertical axis: Inverse of susceptibility as a function of temperature. The red dashed line represents a linear regression of the high– temperature regime. (c) The M(H) for CCPS sample under H//ab (upper panel) and H⊥ab (lower panel) magnetic fields from T = 300 to 5 K. Inset: The derivative dM/dH for H//ab at T = 5 K at a field range of 0 to 9 T. The red and blue arrows denote $H_{SF}$ and $H_{FM}$, respectively. (d) Zoom–in data of in–plane M(H) below $T_N$ to show the spin–flop transition (red arrow). Figs. 3c and 3d have the same color codes. (e) The magnetic phase diagram constructed from the in–plane M(H) measurement presented in Fig. 3c.*

In addition to the spin reorientation phenomenon, the absolute magnitude of magnetization in Cr–based compounds also present a notable point of interest. Studies based on magnetometry [11] and X–ray magnetic circular dichroism (XMCD) [50] have mainly attributed the net magnetic moment of such compounds to the $Cr^{3+}$ spin, with a minimum contribution from orbital moment [11,50]. Similar magnetic behavior can also be seen in the magnetization measurements of CCPS. At $T$ = 5 K, the saturation magnetization for both field directions attains a value of ~2.94 $\mu_B$ per f.u., which is in good agreement with the spin–only moment of $g_S S \mu_B \approx 3$ $\mu_B$, where $g_S$ ($\approx 2$) is the free electron Landé $g$-factor and S = 3/2 for a $Cr^{3+}$ ion. The Curie–Weiss analysis of susceptibility yields a $\mu_{eff}$ = (3.86±0.03) $\mu_B$. This value aligns well with the theoretical result of $(\mu_{eff})_{theo} = g_S\sqrt{[S(S+1)]}\mu_B$ = 3.87 $\mu_B$. These data suggest that the magnetic moment in CCPS originates predominantly from the spin–only contribution, implying a negligible orbital magnetic moment for a $Cr^{3+}$ ion. This distinctive characteristic of $Cr^{3+}$ ions placed in an octahedral crystal field is attributable to orbital moment quenching arising from their high–spin $t^3_{2g}$ electronic configuration. Accordingly, these measurements serve primarily to benchmark our samples and establish a consistent framework for examining the $Cr^{3+}$ spin dynamics with minimal orbital contributions.

Probing the magnetization dynamics may provide deeper insight into the unique magnetic behavior of $Cr^{3+}$ spins. Here, we examine the AFM dynamics of CCPS using AFMR measurements performed across a range of temperatures and frequencies. The schematic in Fig. 4a depicts the Larmor precession and the fundamental concept behind the magnetic resonance phenomenon. The net magnetic moment of the spin system, given by the magnetization vector $M$ (purple arrow), undergoes precessional motion around the applied static magnetic field $H$ (red arrow), which is aligned perpendicular to the microwave magnetic field $H_{RF}$ (orange double arrow). When the Larmor precession frequency of $M$ matches that of the $H_{RF}$, resonance occurs, and the system absorbs microwave energy, resulting in a distinct change in microwave transmission or reflection. The microwave absorption in AFMR measurement is detected by scanning the DC magnetic field at a fixed excitation frequency. Owing to two anti–parallel magnetic sublattices in antiferromagnets, their precession dynamics involve coupled modes, and the subsequent resonance is relatively complex compared to the typical single–mode dynamics present in FMR. The nature of the AFMR in CCPS can be understood by studying the evolution of resonance as a function of input frequencies and modeling it with the coupled Landau–Lifshitz–Gilbert (LLG) equation, using a macrospin approximation for each spin sublattice [51]. This model assumes the uniform magnetization direction in two oppositely aligned magnetic sublattices 1 and 2, each with a magnitude of saturation magnetization $M_S$, which are denoted by the unit vectors $\hat{m}_1$ and $\hat{m}_2$, respectively, as shown in the inset of Fig. 3a. The exchange forces are treated as molecular fields $-H_E\hat{m}_2$ and $-H_E\hat{m}_1$ acting on the sublattices 1 and 2, respectively, where $H_E$ is the interlayer exchange field. The coupled LLG equation, excluding damping, is given below:

$$\frac{d\hat{m}_{1(2)}}{dt} = -\mu_0\gamma\hat{m}_{1(2)} \times [H - H_E\hat{m}_{2(1)} - M_s(\hat{m}_{1(2)} \cdot \hat{z})\hat{z}] + \tau_{1(2)}, \quad (1)$$

where $\mu_0$ is the permeability constant, $\gamma$ is the gyromagnetic ratio, and $\hat{z}$ is the direction perpendicular to the sample plane, i.e., along the crystallographic $c$–axis. $\tau_1$ and $\tau_1$ are the torques due to the microwave field on sublattices 1 and 2, respectively. When the magnetic field $H$ is applied along the $ab$–plane, the above equation is symmetric under the combined symmetry operation of twofold $C_2$ rotation and sublattice exchange [28,35]. The two independent modes, namely optical and acoustic modes with opposite parity, emerge under such symmetry. Depending on the orientation of $H$ and $H_{RF}$ fields, these two modes co–exist or appear selectively in the resonance due to their different symmetry and coupling conditions. Prior work on the related

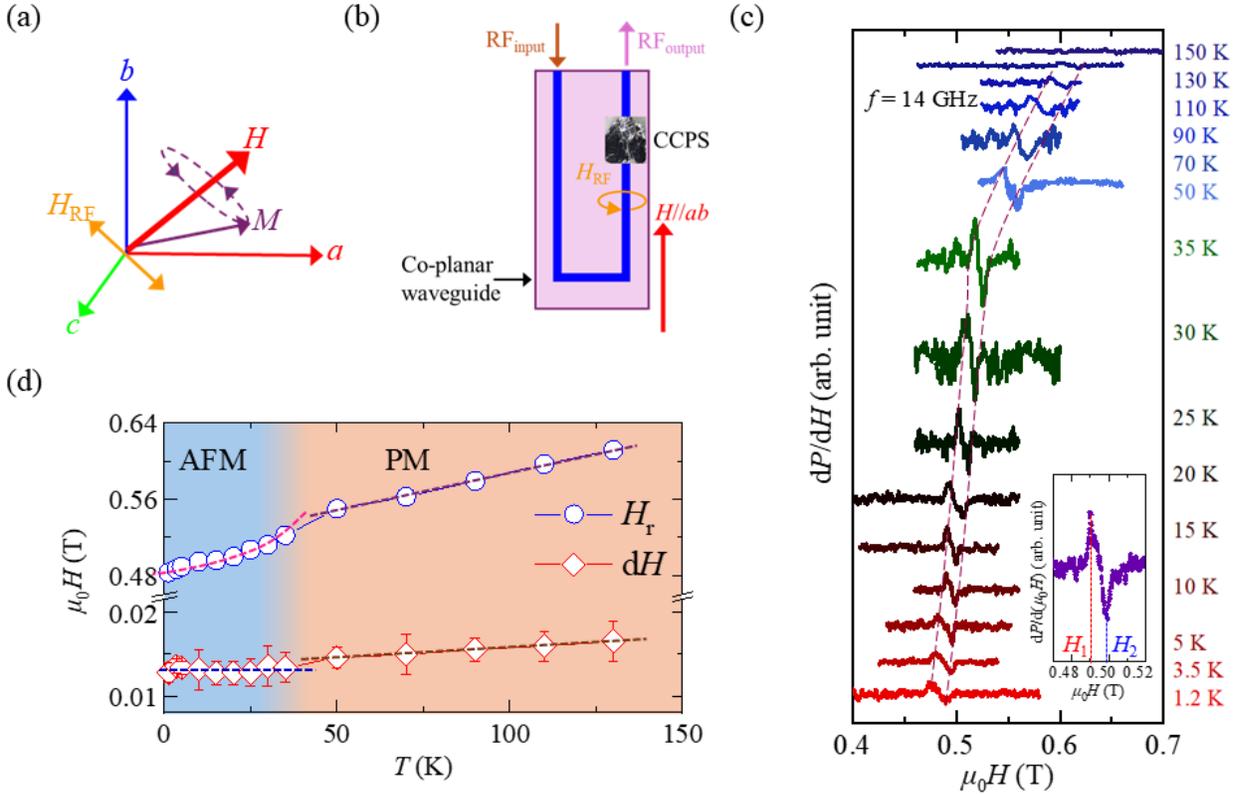

*Figure 4.* (a) A sketch of Larmor precession of magnetization (M) and orthogonality of the rf ($H_{RF}$) and dc (H) fields leading to magnetic resonance. Here, a, b, and c represent the crystallographic axes. (b) The grounded co–planar waveguide on which the crystal is glued with Apizone–N grease such that the ab–plane of the crystal is on the waveguide, and both H and $H_{RF}$ fields lie within the ab–plane of a CCPS crystal. (c) The microwave absorption spectra of the CCPS single crystal,

*recorded at f = 14 GHz over a temperature range of 1.2 to 130 K. Inset of Fig. (c) shows the method to extract the resonance field ($H_r$) and peak-to-peak linewidth (dH) of resonance. (d) Temperature dependence of $H_r$ and dH across the PM and AFM phases. The dashed lines are a guide to the eye.*

compound $CrCl_3$ has resolved both optical and acoustic modes of AFMR in the GHz frequency range when the $H$ and $H_{RF}$ fields are parallel [28]. In contrast, only the acoustic mode persisted under a perpendicular field configuration [28].

The experimental setup for the AFMR spectroscopy used here is illustrated in Fig. 4b. Both $H$ and $H_{RF}$ fields lie within the *ab*–plane of a CCPS crystal, yet oriented perpendicular to each other. As anticipated, a single resonance is detected for mutually perpendicular $H$ and $H_{RF}$ fields across the entire frequency range $f = 4 –14$ GHz within the temperature interval $T = 1.2$ to 130 K. The typical absorption spectra recorded at $f = 14$ GHz and shown in Fig. 4c, reveal a single resonance mode persisting throughout the temperature range. The resonance field ($H_r$) and peak–to–peak linewidth (dH) of resonance are extracted following a standard method employed in earlier works [12,29,32]. The $H_r$ has been defined as the midpoint between the peak ($H_1$; red dashed line) and valley ($H_2$; blue dashed line) positions of the field derivative of the absorbed power (dP/dH), as illustrated in the inset of Fig. 3c. Furthermore, the dH is given by dH = ($H_2 – H_1$), reflecting the full width at half maximum of the Lorentzian absorption signal. The temperature dependence of $H_r$ and dH is summarized in Fig. 4d. The resonance becomes undetectable on further heating above $T > 130$ K. The suppression of $H_r$ on cooling under $H//ab$ field supports the in–plane magnetic easy axis in CCPS, consistent with the typical temperature–dependent $H_r$ trend seen in other Cr-based antiferromagnets [27,30] and ferromagnets [29,52]. Notably, a pronounced change of slope of the $H_r$ vs $T$ plot occurs near the PM to AFM phase boundary. The $H_r$ reduction continues from $\mu_0 H_r \approx 0.52$ T at 35 K to $\mu_0 H_r \approx 0.48$ T at 1.2 K. However, the change is not strictly linear with temperature below $T_N$ (denoted by the dashed lines in Figs. 4c and d), suggesting the modulation of spin dynamics due to the onset of AFM ordering. Additionally, the dH also exhibits a distinct temperature dependence across the PM and AFM phases. The dH reduces linearly with decreasing temperature above $T_N$ but displays a weaker temperature dependence upon entering the AFM phase (Fig. 3d). Indeed, contrasting temperature dependence of dH across the magnetic ordering temperature is commonly seen in other Cr compounds [26,30,52] as well. The microwave absorption below $T_N$ in CCPS, where $H_r$ is notably smaller than the $H_{FM}$ field derived from M(H) measurements (Fig. 3c), is likely associated with AFMR caused by its weak interlayer AFM coupling [28,35]. However, the persistence of resonance above $T_N$ in these vdW materials is surprising. Previous studies have attributed this behavior either to PM resonance originating from localized $Cr^{3+}$ ions [28] or to the presence of short–range magnetic correlation above the ordering temperature [27,30]. As summarized in Table II, the observation of a well–defined FMR/ESR signal extending to temperatures several times higher than $T_N$ or $T_C$ is a generic feature of Cr–based vdW magnets, with $T_{SR}/(T_N$ or $T_C)$ ranging from ~2 to ~12. Importantly, the resonance at $T_{SR}$ remains relatively narrow, with d$H(T_{SR})$ of only ~$10^{-1}$–$10^{-2}$ T for most compounds, which points to robust short–range magnetic correlations that survive deep into the PM regime.

*Table II. Summary of FMR/ ESR parameters for representative Cr–based AFM and FM compounds. Here, $T_{SR}$ denotes the characteristic temperature up to which the FMR/ESR signal persists above the Néel temperature ($T_N$) or Curie temperature ($T_C$), reflecting the presence of short–range magnetic correlations. The ratio $T_{SR}/(T_N$ or $T_C)$ quantifies the relative extent of this correlated regime, while d$H(T_{SR})$ represents the FMR/ESR linewidth evaluated at $T_{SR}$.*

| Materials | $T_{SR}$ (K) | $T_{SR}/(T_N$ or $T_C)$ | d$H(T_{SR})$ (T) | Ref. |
|---|---|---|---|---|
| $CuCrP_2S_6$ | 130 | ~4.1 | ~0.016 | This work |
| $CrCl_3$ | 175 | ~12.5 | – | [27] |
| $CrBr_3$ | 300 | ~8.1 | ~0.03 | [52] |
| $CrGeTe_3$ | 298 | ~4.4 | ~0.45 | [29] |
| CrSBr | 298 | ~2.2 | ~0.03 | [30] |

Deeper insights into the AFM dynamics of CCPS can be achieved by analyzing the variation of $H_r$ with applied microwave frequency $f$. The $f(H_r)$ function extracted from AFMR measurements at various temperatures is presented in Fig. 5a. As described by the LLG equation (eq. 1), this behavior can be expressed by the following relation [28,51]:

$f = \mu_0 \gamma' [\sqrt{\{2H_E(2H_E + M_s)\}}] H_r / 2H_E$,   (2)

where $\gamma'$ is the reduced gyromagnetic ratio ($\gamma' = |\gamma|/2\pi$). To verify the extent of orbital moment quenching and confirm that the net magnetic moment originates primarily from the spin–only value of the $Cr^{3+}$ ion, the $\gamma'$ below $T_N$ is evaluated by fitting $f$ as a function of $H_r$ using eq. 2. This quantity allows for the calculation of the Landé $g$-factor ($g$) via the relation $g = |\gamma|\hbar/\mu_B$. On the other side, in the PM regime, where the magnetization is proportional to the applied field, i.e., $M = \chi(T)H$, the resonant frequency is given by [28]:

$$f = \mu_0 \gamma' [\sqrt{\{1 + \chi(T)\}}]H, \quad (3)$$

which is employed to fit the $f$ vs $H_r$ data above $T_N$. These fittings yielded a value of $\gamma'$ between $\sim(28.80\pm0.13)$ GHz/T and $\sim(27.56\pm0.26)$ GHz/T within the temperature range of $T = 1.2$ to 130 K. A comparable value of $\gamma' \approx 28$ GHz/T has been reported for $CrCl_3$ [28,32], implying a similar AFMR dynamics for the $Cr^{3+}$ ion. We estimated the $g$–value using $\gamma'$ and plotted its evolution with temperature in Fig. 5b. Interestingly, the $g$–factor exhibits a weak temperature dependence and attains a value between ~(2.06±0.01) and ~(1.97±0.02) across AFM and PM phases, which is close to the spin–only Landé g-factor $g_s \approx 2$. This indicates a negligible orbital contribution to the total magnetic moment, in agreement with the saturation magnetization and effective moment derived from magnetization measurements, which are close to the theoretical spin–only value for the high–spin $Cr^{3+}$ ion (S = 3/2). Using similar AFMR spectroscopy, a comparable g–factor close to $g_s \approx 2$, irrespective of the crystallographic orientations, has been determined for related AFM compounds such as $CrCl_3$ [32] ($g \approx 1.989$ for H//ab and $g \approx 1.982$ for H⊥ab) and CrSBr [30] ($g \approx 1.996±0.001$, 2.003±0.001, and 1.966±0.001 for H//a, H//b, and H//c, respectively). Meanwhile, significantly larger g–factors, i.e., $g = 2.18±0.02$ for H//ab and $g = 2.10±0.01$ for H⊥ab have been estimated for a Cr–based ferromagnet $Cr_2Ge_2Te_6$ using FMR measurement [31]. The deviation from the spin–only g–factor in $Cr_2Ge_2Te_6$ suggests a finite orbital contribution to the total magnetization. An orbital moment of 8% and 5% along the in-plane and out–of–plane directions, respectively, has been confirmed in this compound [31]. Such a non–negligible orbital contribution strengthens the SOC effects and consequently amplifies the magnetic anisotropy of the system. In fact, a sizeable magnetocrystalline anisotropy energy of $(0.48±0.02)\times10^6$ erg/cm³ (i.e., ~0.082±0.0034) meV per formula unit has been reported for $Cr_2Ge_2Te_6$ [29]. Additionally, a similar FMR study on a related ferromagnet, $CrI_3$, has also revealed a strong magnetic anisotropy originating from the anisotropic Kitaev interactions [24]. This pronounced anisotropy accounts for the direction–dependent g–factors observed in ferromagnetic $Cr_2Ge_2Te_6$ [31], in stark contrast to the nearly isotropic g–values reported for its AFM counterparts [30,32]. These observations highlight the distinct microscopic mechanisms that govern the AFM and FM orderings in Cr–based magnetic materials.

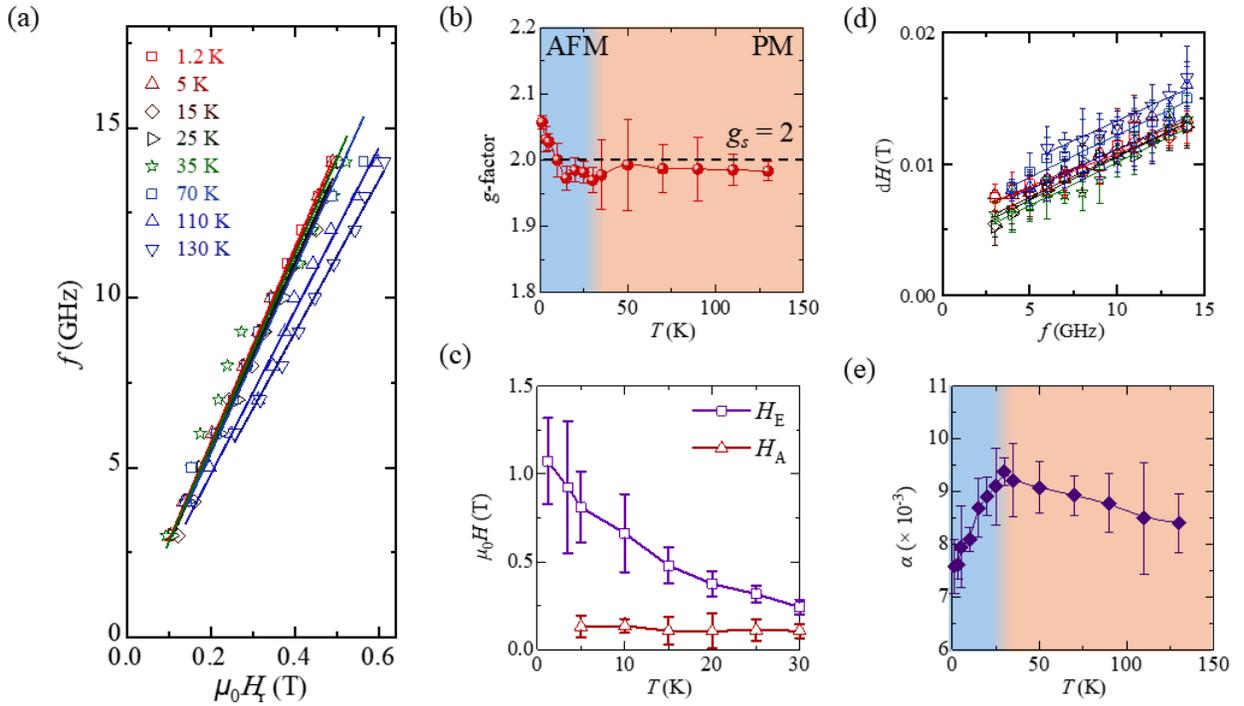

*Figure 5.* (a) Input frequency (f) as a function of resonance field $H_r$ plotted at several temperatures in the range of 1.2 to 130 K for a CCPS crystal with its ab–plane parallel to the directions of the dc and RF fields. The solid lines indicate fitting of f vs $H_r$ to a function presented in equation 2 (main text). (b) Temperature dependence of the Lande g–factor (g) across the PM and AFM states. The dashed line represents the spin–only g–factor $g_s = 2$. (c) Temperature dependence of the exchange ($H_E$) and effective anisotropy ($H_A$) fields below $T_N$. (d) The linewidth (dH) as a function of f over a temperature range of 1.2 to 130 K. The solid lines indicate fitting of dH vs f to a function presented in equation 4 (main text). The figures 5a and 5d use the same color codes. (e) Temperature dependence of the Gilbert damping parameter (α). The different colored regions in Figs. 5b and 5e represent AFM (blue) and PM (orange) states.

While the ferromagnets such as $Cr_2Ge_2Te_6$ [31] and $CrI_3$ [24] exhibit strong magnetic anisotropy, the extent to which this effect shapes the Cr–based AFM systems remains less clear. To address this, we determined the AFM exchange field $H_E$ and the effective anisotropy field $H_A$ for CCPS, thereby clarifying the relative roles of exchange interactions and anisotropy in Cr antiferromagnets. The $H_E$ field is directly extracted from the f vs $H_r$ fitting using eq. 2 (Fig. 5a). While analogous AFMR approaches have been successfully used to determine $H_E$ in layered antiferromagnets such as $CrCl_3$ [28] and CrSBr [35], a direct experimental determination of $H_E$ for CCPS has been notably absent to date. With this $H_E$ parameter, the $H_A$ field is calculated through the expression:

$$H_{SF} = \sqrt{2H_E H_A - H_A^2}, \quad (3)$$

where the spin-flop field $H_{SF}$ in the temperature range of 5 to 30 K is derived from $M$(H) measurements (Fig. 3c). As seen in Fig. 5c, the $H_E$ enhances as the temperature is lowered, indicating a steady strengthening of exchange interactions due to reduced thermal fluctuations. The $H_A$ likewise grows as the system is cooled, but its temperature dependence is weak, and the magnitude remains far below that of $H_E$ across the full temperature interval. The much smaller $H_A$ relative to $H_E$ corresponds well with the orbital moment quenching and the resulting weak magnetic anisotropy for Cr–based antiferromagnets, as discussed earlier. Further, within the standard two–sublattice AFMR framework, the resonance frequency can be approximated as $\omega \propto \gamma\sqrt{(H_E H_A)}$ [35]; therefore, the small anisotropy field $H_A$ also naturally accounts for the comparatively low AFMR frequencies observed in CCPS as well as in $CrCl_3$ [28], despite a sizable exchange field $H_E$. However, the stronger $H_E$ field in CCPS leads to higher resonance fields $H_r$ at similar frequencies relative to the $H_r$ of $CrCl_3$ [28]. In such cases where magnetic anisotropy is negligible, the $H_E$ also becomes the dominant factor governing magnetic order; consequently, its magnitude scales directly with the ordering temperature. For instance, a considerably larger $\mu_0 H_E \approx (1.07 \pm 0.24)$ T at 1.2 K is estimated for CCPS ($T_N \approx 32$ K) compared to a value of $\mu_0 H_E \approx 0.105$ T measured in $CrCl_3$ with $T_N \approx 14$ K [28]. Furthermore, the exchange field for CCPS (~1 T) is significantly smaller than the value of $\mu_0 H_E \approx 100$ T calculated in a pioneering work by Kittel [53] for another Cr antiferromagnet, $Cr_2O_3$, which has a much higher $T_N$ (~314 K) [54] than CCPS. In addition to the ordering temperature, the spin reorientation and the resulting field–induced SF transition in CCPS are likely governed by exchange interactions rather than anisotropy. Fig. 3d shows that the $H_{SF}$ systematically increases upon cooling and mirrors the behavior of $H_E$, suggesting that exchange interactions primarily drive the SF transition in CCPS. This underscores the dominant role of isotropic Cr–Cr exchange interactions in stabilizing the AFM order in CCPS, with magnetic anisotropy playing only a marginal role.

In 2D honeycomb AFM systems with weak SOC, such as CCPS, the damping primarily arises from the relaxation of the electron spin via the exchange field [55]. This relaxation process is quantified by the phenomenological Gilbert damping parameter ($\alpha$), which is extracted by linearly fitting d$H$ vs $f$ curves shown in Fig. 4d using the following equation

$$dH = (4\pi/\gamma)\alpha f + dH_0, \quad (4)$$

where d$H_0$ represents the inhomogeneous linewidth contribution and the $\gamma$ is the gyromagnetic ratio that was already evaluated by fitting $f$ as a function of $H_r$ using eq. 2. The temperature dependence of the $\alpha$ is plotted in Fig. 5e. For CCPS, the $\alpha$ ranges between $\approx (7.5$ to $9.4) \times 10^{-3}$ over the temperature interval of 1.2 to 130 K. Notably, this $\alpha$ is comparable to other Cr compounds, $Cr_2Ge_2Te_6$ [56,57] and $CrBr_3$ [9]. Such ultra-low magnetic damping is attributed to the weak SOC originating from the quenched orbital moment of the $Cr^{3+}$ ion ($g \approx 2$), which suppresses the spin–phonon scattering and subsequently reduces the transverse magnetization relaxation [9,56,57]. Furthermore, because electrons are localized in an insulating system like CCPS, spin relaxation via electron–electron scattering is also weaker compared to that in conductive systems [9,56,57]. This highlights the robustness of spin coherence in such vdW semiconductors. Furthermore, the evolution of $\alpha$ with temperature is also intriguing; it increases upon heating from 1.2 K until reaching a maximum value near the PM–AFM phase boundary (Fig. 5e). This behavior can be ascribed to enhanced spin fluctuations promoting damping in the vicinity of the magnetic phase transition [58]. Further increasing the temperature reduces $\alpha$ before vanishing above $T = 130$ K. The persistence of damping even after the long-range magnetic order vanishes above $T_N$ is intriguing. As mentioned earlier, the short–range magnetic correlations well beyond $T_N$ [27,30] may lead to correlated spin clusters that can support collective spin excitations [59,60], giving rise to spin relaxation and finite damping even in the PM phase.

The dominant role of Cr–Cr spin interactions is further highlighted by the characteristic field-driven AFM–FM transition commonly observed in Cr–based antiferromagnets. As discussed earlier, these materials display a generic feature of field–induced FM–like saturation (Fig. 1b), which is also evident in the $M(H)$ measurements of CCPS (Fig. 3c). Although CCPS exhibits weak magnetic anisotropy, the isothermal magnetization still shows anisotropic features, manifested by the spin–flop transition only along the $ab$-plane and the direction–dependent saturation fields (Fig. 3c). However, upon full saturation, the magnetization becomes isotropic, reflecting the underlying isotropic Cr–Cr exchange interactions that govern the behavior of $Cr^{3+}$ moments. Such nearly isotropic magnetic behavior is likewise reflected in the high–field magnetic resonance response. Applying the field $H > 2H_E$ drives the system into the FM state [28], where the LLG equation (eq. 2) is transformed to the Kittel formula, $f = \mu_0 \gamma' [\sqrt{\{H(H + M_s)\}}]$ and the associated AFMR evolves into a uniform FMR mode characteristic of ferromagnets. For CCPS, the $H_E$ is close to ~1T, implying that FMR can be accessed for applied fields $\mu_0 H_E > 2$ T. However, magnetic resonance spectroscopy could not be performed at such high fields due to the limited frequency range of our cryo–FMR spectrometer [38]. To overcome this, we carried out complementary high–frequency

FMR measurements at $f = 240$ GHz across a temperature range $T = 5–261$ K (Fig. 6a) under $H // ab$ field configuration using the high–frequency EPR facilities at the NHMFL. To date, high–frequency FMR investigations have focused primarily on Cr–based vdW FM compounds such as CrBr$_3$ [26] and Cr$_2$Ge$_2$Te$_6$ [29] with only limited reports on related antiferromagnets such as CrCl$_3$ [27], and none to our knowledge for CCPS. At high temperatures, a single PM signal with a Lorentzian line shape is observed in CCPS (Fig. 6a). As the system approaches $T_N$, this signal progressively splits into multiple features. At such high frequencies, the resonance spectra correspond to FMR. Therefore, the low–temperature regime below $T_N$ can be regarded as a high–frequency (or high–field) ferromagnetic state, hereafter denoted as FM (Fig. 6a). The multiple peaks in the high–frequency (240 GHz) FMR signal are also observed in CrBr$_3$ [26], which can be caused by several factors, including magnetocrystalline anisotropy, magnetic inhomogeneity, and magnetic phase separation [26]. Furthermore, strong absorption near resonance induces a rapid frequency dependence of the refractive index ($n$), causing symmetric phase shifts and interference effects. For typical dielectrics or semiconductors with $n \approx 2$ at microwave frequencies, the half–wavelength ($\lambda/2$) inside the material for a 240 GHz wave is approximately 300 μm, comparable to the sample thickness and thus enhancing these effects. Finite–size and surface deviations from an ideal thin plate further complicate the spectral structure. The temperature dependence of $H_r$ is summarized in Fig. 6a (inset), which decreases on lowering the temperature, consistent with the AFMR results in Fig. 3d, and confirms the in-plane magnetic easy axis in CCPS. This easy–axis alignment along the ab–plane is further verified by angle–dependent FMR measurements at $T = 10$ K. As shown in Fig. 5b, FMR spectra were recorded for field orientations varying from θ = 0° (out–of–plane) to θ = 90° (in–plane), with the measurement geometry illustrated in the inset. With increasing angle θ, $H_r$ systematically shifts to lower values, reaching its maximum in the out–of–plane configuration and minimum in the in–plane direction, in excellent agreement with the in–plane moment orientation in CCPS. Furthermore, the $H_r$ as a function of θ angle is fitted using the model: $H_r(\theta) = A(3\cos^2\theta - 1) + D$ (red solid line. Such $(3\cos^2\theta - 1)$ dependence of $H_r$ is a characteristic signature of 2D magnetic systems and has been reported in high–frequency FMR studies of CrBr$_3$ [26] and Cr$_2$Ge$_2$Te$_6$ [29]. This behavior further corroborates the AFMR and FMR arising from the quasi–2D magnetic dynamics of CCPS. Assuming the demagnetizing factors for an infinitely thin platelet ($N_x = N_y = 0$ and $N_z = 1$), the demagnetizing field ($H_d$) at the Cr sites is estimated to be ~1.43 T from the FMR relation: $f = (g\mu_B/h)(H_r - H_d)$, using an applied frequency of f = 240 GHz, the out–of–plane resonance field $\mu_0 H_r \approx 10$ T (Fig. 6b), and $g \approx 2$. For a uniform CCPS sample, the $H_d$ should be equal to its volume magnetization (Magnetic moment/Unit cell volume) ≈ 0.12 T, using $\mu_{eff} \approx 3.86\ \mu_B$ obtained from CW fitting (Fig. 3b) and unit cell volume ≈ 782 Å$^3$ [11]. However, the substantially larger $H_d$ field observed here indicates enhanced local dipolar interactions arising from the layered structure, consistent with the quasi–2D magnetic nature inferred from the FMR and AFMR results.

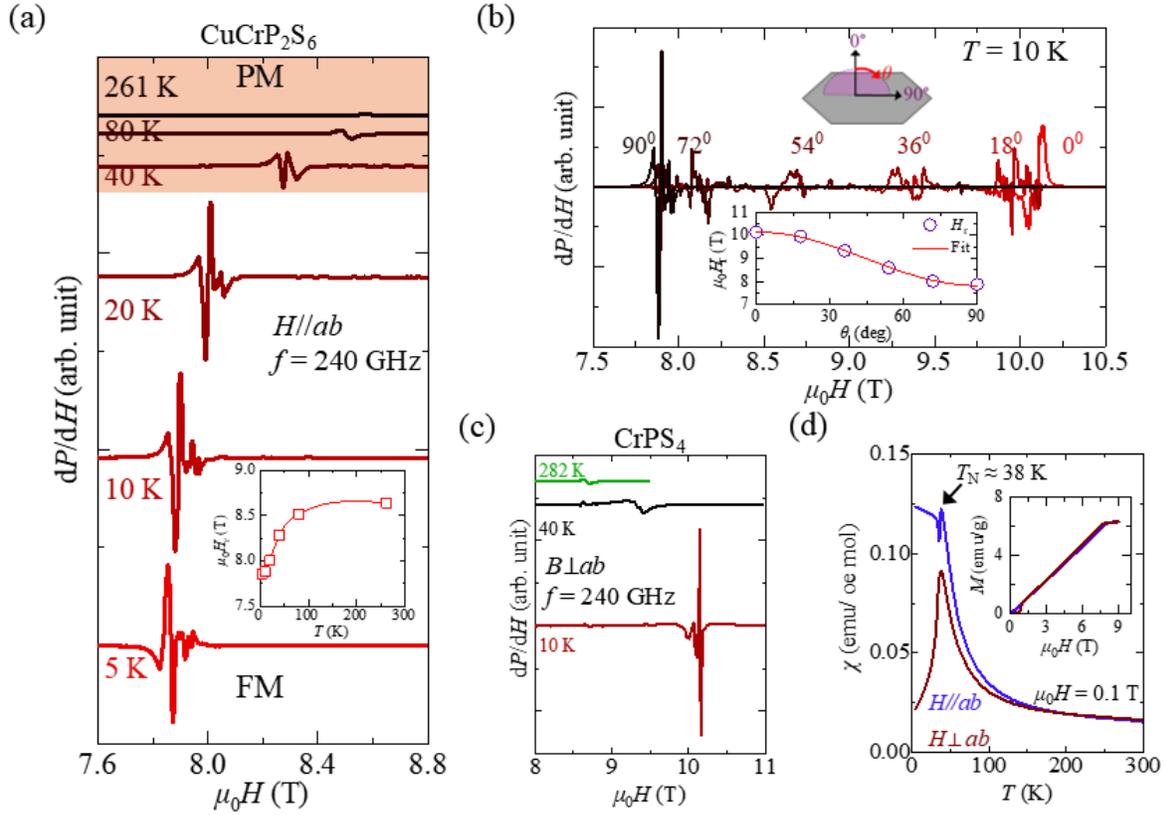

***Figure 6.*** *(a) The FMR spectra of a CCPS single crystal, recorded at 240 GHz over a temperature range of 5 to 261 K under the H//ab applied field. Inset: Temperature dependence of $H_r$. (b) The FMR spectra of the CCPS crystal at 240 GHz and 10 K under various field orientations from out–of–plane ($\theta = 0°$) to in–plane ($\theta = 90°$) directions. Inset: Angular dependence of $H_r$, which is fitted by the equation: $H_r(\theta) = A(3\cos^2\theta - 1) + D$ (red solid line). (c) The FMR spectra of the CrPS$_4$ single crystal, recorded at 240 GHz and $T = 10$, 40, and 282 K under the $H \perp ab$ field. (d) Temperature dependence of susceptibility ($\chi$) for the CrPS$_4$ sample under $H||ab$ and $H \perp ab$ magnetic fields of $\mu_0 H = 0.1$ T. The black arrow denotes the $T_N \approx 38$ K. Inset: M(H) for the CrPS$_4$ sample under H//ab and $H \perp ab$ magnetic fields from $T = 10$ K.*

The resonance fields in CCPS span from $\mu_0 H_r \approx 8$ T along the magnetic easy axis to ~10 T along the hard axis at $T = 10$ K (Fig. 5b). Furthermore, $\mu_0 H_r \approx 8$ T is consistently observed below $T_N$ in the temperature interval $T = 5–20$ K (Fig. 5a). Comparable resonance fields at similar frequencies have been reported in related Cr–based systems, including $CrCl_3$ [27], $CrBr_3$ [26], and $Cr_2Ge_2Te_6$ [29]. Taken together, these results point to a generic physical origin that governs the field–driven AFM–to–FM transition in Cr antiferromagnets. To further clarify whether Cr–Cr exchange interactions are the underlying mechanism, we also carried out high–frequency FMR measurements at $T = 10$, 40, and 282 K on a related antiferromagnet, $CrPS_4$ (Fig. 6c). This compound was chosen because the $Cr^{3+}$ ions reside in an approximately $O_h$ symmetric environment (Table I) and the absence of heavier ligands minimizes additional sources of magnetic anisotropy. In other words, $CrPS_4$ most likely captures the intrinsic magnetic behavior of $Cr^{3+}$ moments and thus provides a meaningful comparison to CCPS. The magnetization measurements reveal $T_N \approx 38$ K for $CrPS_4$ (Fig. 6d), consistent with earlier works [14,22]. As the system is cooled below $T_N$, $\chi_\perp$ drops sharply in comparison to $\chi_{//}$, indicating the out–of–plane easy axis in this compound. This is confirmed by the metamagnetic SF transition in out–of–plane magnetization, accompanied by its smaller saturation field than that of in–plane magnetization (inset; Fig. 6d). Therefore, the FMR spectra for this compound were collected under the $H \perp ab$ field. The high–temperature resonance measurement reveals a single PM Lorentzian signal, while, as expected, multiple features emerge at $T = 10$ K (Fig. 6c). Notably, we obtain $\mu_0 H_r \approx 10.2$ T for the out–of–plane direction, which is in excellent agreement with the corresponding value observed in CCPS (Fig. 6b). The identical FMR behavior observed in two distinct Cr–based antiferromagnets suggests similar local dipolar interactions, further underscoring the universality of Cr–Cr exchange interactions as the fundamental microscopic mechanism that dictates AFM order and drives the field-induced AFM–FM transition in vdW Cr compounds.

In conclusion, we have investigated the $Cr^{3+}$ spin dynamics of the vdW antiferromagnet $CuCrP_2S_6$ through magnetization, broadband AFMR, and high–frequency FMR measurements. Our study reveals that the magnetic behavior of CCPS is governed predominantly by isotropic Cr–Cr exchange interactions, while orbital moment quenching minimizes the influence of magnetic anisotropy. These fundamental interactions dictate the magnetic order, spin reorientation, and magnetic damping, while enabling field–induced FM–like polarization. A key result of this investigation is the establishment of the universality of the magnetic behavior of several $Cr^{3+}$ ion–based vdW AFM compounds. Collectively, CCPS emerges as a prototypical vdW antiferromagnet with microwave–tunable AFM dynamics, ultra–low damping, and possible room–temperature ferroelectricity, offering a promising platform for next–generation spintronic applications. Future investigations of a possible coupling between AFM and antiferroelectric order parameters may help elucidate potential magnetoelectric correlations and guide the design of multifunctional vdW materials.

**Acknowledgements**

This work at Morgan State University has been funded by the Department of Defense through grant # W911NF2120213. The contribution of J. V. Tol was funded by the NHMFL and National Science Foundation Cooperative Agreement No. DMR-2128556* and the State of Florida.